# LoRa Backscatter: Enabling The Vision of Ubiquitous Connectivity

VAMSI TALLA[†], MEHRDAD HESSAR[†], BRYCE KELLOGG, ALI NAJAFI, JOSHUA R. SMITH
AND SHYAMNATH GOLLAKOTA, University of Washington
[†]Co-primary Student Authors

The vision of embedding connectivity into billions of everyday objects runs into the reality of existing communication technologies — there is no existing wireless technology that can provide reliable and long-range communication at tens of microwatts of power as well as cost less than a dime. While backscatter is low-power and low-cost, it is known to be limited to short ranges. This paper overturns this conventional wisdom about backscatter and presents the first wide-area backscatter system. Our design can successfully backscatter from any location between an RF source and receiver, separated by 475 m, while being compatible with commodity LoRa hardware. Further, when our backscatter device is co-located with the RF source, the receiver can be as far as 2.8 km away. We deploy our system in a 4,800 $ft^2$ (446 $m^2$) house spread across three floors, a 13,024 $ft^2$ (1210 $m^2$) office area covering 41 rooms, as well as a one-acre (4046 $m^2$) vegetable farm and show that we can achieve reliable coverage, using only a single RF source and receiver. We also build a contact lens prototype as well as a flexible epidermal patch device attached to the human skin. We show that these devices can reliably backscatter data across a 3,328 $ft^2$ (309 $m^2$) room. Finally, we present a design sketch of a LoRa backscatter IC that shows that it costs less than a dime at scale and consumes only 9.25 $\mu$W of power, which is more than 1000x lower power than LoRa radio chipsets.

## 1 INTRODUCTION

Embedding cheap connectivity into billions of everyday objects has been a long standing vision in the ubiquitous computing community. There is however a significant disconnect between the ubiquity articulated by this vision and the capabilities of today's communication technologies. Active radio technologies including Wi-Fi, ZigBee, SigFox [10], LoRa [18] and LTE-M [9] provide reliable coverage and long ranges but are power consuming and cost at least 4–6 dollars [12, 21]; making them too expensive for embedding into objects at scale. Further, given their high peak current and power requirements, active radios significantly deteriorate battery life and are incompatible with emerging small, flexible and ultra thin printed batteries [4, 7, 8] that promise innovative applications across healthcare, wearable devices and cosmetics [25, 42, 43].

Backscatter promises to be an extremely low power, smaller and cheaper alternative to active radios. Given the absence of expensive radio analog components including RF oscillators, decoupling capacitors and crystals, backscatter designs including passive RFID and Wi-Fi backscatter [34, 36] cost only a few cents to manufacture at scale [17]. Further, they consume three to four orders of magnitude lower power and peak current than radios and hence can operate with emerging flexible, printed and ultra thin printed battery technologies. However, despite all these benefits, current backscatter designs have seen very limited adoption beyond RFID applications. This is because current backscatter designs are unreliable, limited in operating range, and in fact today cannot achieve robust coverage across rooms [2, 30, 34, 59] as outlined in Fig. 2 and Table 1 [1] . Furthermore, since RF signals get significantly attenuated by the human body, backscatter range is limited to tens of centimeters in a number of healthcare and wearable device applications [34].

---

[1]We note that active RFID does not use backscatter. Instead, it uses radios, is power consuming and costs between 15–100 dollars [27, 57].

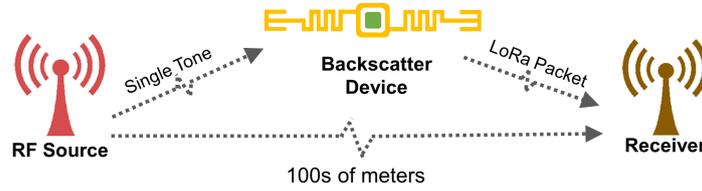

Fig. 1. **LoRa backscatter deployment.** The LoRa Backscatter device consumes 9.25 μW, operates at 100s of meters and can be powered by flexible printed batteries and button cells (10 cents), a capability that cannot be achieved with radios. The RF source transmits a single tone that the backscatter device uses to synthesize CSS signals. The challenge is that at the receiver, the backscatter signal is not only drowned by noise but also suffers interference from the RF source.

This paper questions the conventional wisdom that backscatter is a *short-range* system. Specifically, we ask if one can achieve wide-area backscatter communication with a range of hundreds of meters, if not kilometers. A positive answer would give us the best of both worlds: long-range reliable communication capabilities of radios at the low-power and cost of backscatter hardware. This enables, for the first time, wide area connectivity for everyday objects and opens up applications in domains like smart cities [13], precision agriculture [14], industrial, medical and whole-home sensing [34], where backscatter is currently infeasible.

To appreciate why this is hard, consider the deployment in Fig. 1. Here the backscatter device reflects signals from an RF source to synthesize data packets that are then decoded by a receiver. The challenge is that, before arriving at the backscatter device, the signals from the RF source are already attenuated. The backscatter device is capable of reflecting these weak signals to synthesize data packets which get further attenuated as they propagate to the receiver. Our experiments show that with a separation of 400 m between the RF source and receiver, the backscattered signal is at -134 dBm. In contrast, the direct signal from the RF source at the receiver is more than a million times stronger at -45 dBm. Thus, the backscatter signal is not only drowned by noise but also suffers significant interference from the RF source.

We present the first wide-area backscatter communication system. Achieving this requires us to satisfy two key constraints. First, the backscatter device should code information in a way that can be decoded at the receiver down to and below -135 dBm signal strength and reliably operate in the presence of strong out-of-band interference. Second, instead of using a custom receiver for the backscattered signal that can be prohibitively expensive (e.g., RFID readers), the backscattered signals should be decoded on readily and cheaply available commodity hardware that would expedite the adoption and development of our design.

To do so, we first profile existing radio technologies in Table. 1, which shows that LoRa provides the highest sensitivity of -149 dBm and supports bit rates of 18 bps to 37.5 kbps, which are sufficient for the majority of IoT applications. Further, LoRa is resilient to both in-band and out-of-band interference [20]. Specifically, the Sx1276 receiver hardware from SEMTECH can reliably decode LoRa packets in the presence of 95 dB higher out of band interference [20]. Motivated by this, we present the design and implementation of the first LoRa backscatter system. At a high level, the RF source transmits a single tone signal, which our backscatter devices use to synthesize LoRa compatible packets. To achieve this, we make two key technical contributions.

● *We introduce the first chirp spread spectrum (CSS) backscatter design.* LoRa uses CSS modulation where, as shown in Fig. 4(a), a '0' bit can be represented as a continuous chirp that increases linearly with frequency, while a '1' bit is a chirp that is cyclically shifted in time. Thus, CSS requires continuously changing the frequency as a function of time, which has not been demonstrated on backscatter hardware. This is challenging since while existing backscatter approaches can generate DBPSK/DQPSK (802.11b/ZigBee [34, 36]) and 2-FSK (Bluetooth [30]) transmissions, they are all limited to discrete digital values. Building on existing radio architectures, we design the first backscatter design in §3.2 that can synthesize continuous frequency modulated chirps, while consuming as low as 9.25 $\mu$W which is 3 orders of magnitude lower power compared to LoRa radios. We also reverse-engineer the proprietary LoRa PHY layer in §3.4 so as to backscatter LoRa-compatible packets.

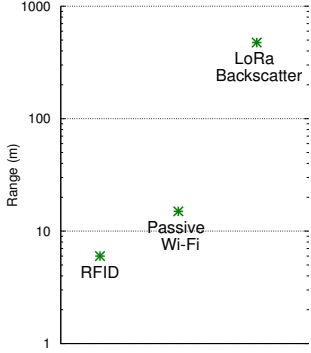

Fig. 2. We compare the operating distance of existing backscatter systems.

| Technology | Sensitivity | Data Rate | Whole Home Coverage | Button Cell | Tiny Solar Cell | Printed Battery |
|---|---|---|---|---|---|---|
| Wi-Fi (802.11 b/g) | -95 dBm | 1-54 Mbps | ✓ | ✗ | ✗ | ✗ |
| LoRa | -149 dBm | 18 kbps–37.5 kbps | ✓ | ✗ | ✗ | ✗ |
| Bluetooth | -97 dBm | 1-2 Mbps | ✗ | ✗ | ✗ | ✗ |
| Sigfox | -126 dBm | 100 bps | ✓ | ✗ | ✗ | ✗ |
| ZigBee | -100 dBm | 250 kbps | ✓ | ✗ | ✗ | ✗ |
| Passive Wi-Fi | -95 dBm | 1-11 Mbps | ✗ | ✓ | ✓ | ✓ |
| RFID | -85 dBm | 40–640 kbps | ✗ | ✓ | ✓ | ✓ |
| LoRa Backscatter | -149 dBm | 18 bps–37.5 kbps | ✓ | ✓ | ✓ | ✓ |

Table 1. **Communication Technologies.** We show the sensitivity and supported data rates for different communication technologies and the feasibility of powering them from different sources.

• *We present the first backscatter harmonic cancellation mechanism.* Backscatter uses a switch to either reflect or absorb the incident RF signals and create square waves. This however creates third and fifth harmonics in adjacent frequency bands when backscattering the signal from the RF source, resulting in interference and affecting network performance. State-of-the-art single-sideband backscatter designs [34] ignore these harmonics and hence create out-of-band interference. Since LoRa has a high sensitivity, this affects other LoRa devices operating in adjacent bands. In §3.3, we present a low-power backscatter design that cancels these sideband harmonics and improves spectral efficiency.

Building on the above techniques, we design a link-layer protocol that enables multiple long range backscatter devices to share the spectrum. We also design a LoRa backscatter IC and estimate the power consumption using Cadence and Synopsis software toolkits [5, 11]. Our results show that our IC design is comparable in area to RFID and consumes as little as 9.25 $\mu$W while generating continuous frequency modulated chirps and performing harmonic cancellation.

Below, we summarize our evaluation in various deployment scenarios.

• We evaluate our design in various line-of-sight scenarios. Our results show that even when the RF source and receiver are separated by 475 m, the backscatter device could operate at all locations between them. Further, when the backscatter device is co-located with the RF source, the LoRa receiver can decode transmissions from as far as 2.8 km from the backscatter device.

• We deploy our system in a one-acre vegetable farm, a 4,800 $ft^2$ (446 $m^2$) house spread across three floors and a 13,024 $ft^2$ (1210 $m^2$) office space spanning 41 rooms separated by wood, concrete walls and metal structures. Our results show reliable backscatter coverage across all these deployments, using only a single RF source and receiver.

• Finally, we show that our design can also enable backscatter in applications that are not favorable for RF propagation. Specifically, we evaluate a contact lens form factor antenna in-vitro and show that it can easily backscatter data from across a 3,328 $ft^2$ (309 $m^2$) room using only a single RF source and receiver. This is orders of magnitude larger than the 35 inch (89 $cm$) range achieved by prior designs [34]. We also build a flexible epidermal patch sensor that backscatters data in the above room, while being attached to human skin.

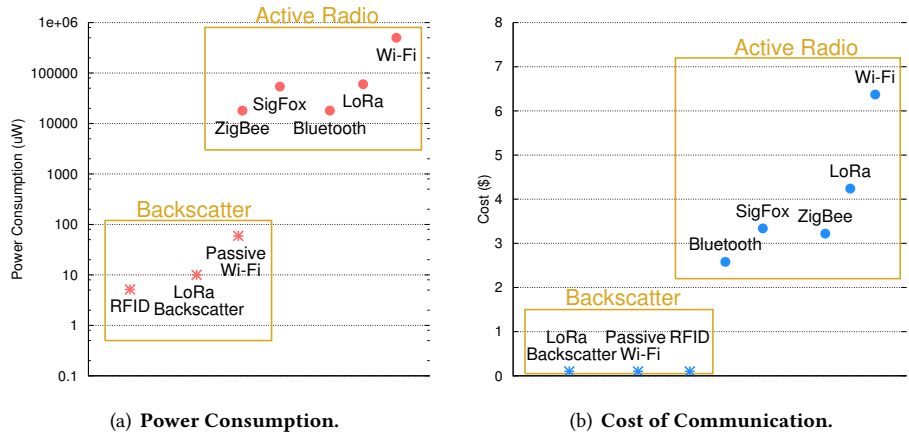

(a) **Power Consumption.**  (b) **Cost of Communication.**

Fig. 3. We compare power consumption and cost of different communication technologies including backscatter and radio techniques.

## 2 THE CASE FOR LORA BACKSCATTER AND RELATED WORK

To demonstrate that LoRa backscatter is the best fit for achieving ubiquitous connectivity, we first compare it with existing solutions and show how it fares across the spectrum of power consumption, cost, size, sources of power and operating range. We then discuss prior work and compare our LoRa backscatter system with state of the art backscatter approaches.

• *Operating range.* Fig. 2 shows the range of different radio and backscatter communication solutions. Radios including Wi-Fi, Bluetooth and ZigBee operate up to 100s of meters while wide area LoRa and SigFox deployments extend operation to kilometers. However, existing backscatter solutions such as RFID and Passive Wi-Fi are limited to tens of meters of operating distance in best-case scenarios. LoRa backscatter instead extends backscatter operation to 100s of meters. This achieves the whole home and office coverage of Wi-Fi and ZigBee radios while delivering the cost, size and power benefits of backscatter described below.

• *Power consumption.* Fig. 3(a) shows the power consumption of popular radio and existing backscatter solutions. It can be seen that radios including Wi-Fi, BLE, ZigBee, Lora and SigFox all consume between 10 to 500 mW, which is 3–4 orders of magnitude higher than the power consumption of backscatter systems including RFID, Passive Wi-Fi and the LoRa backscatter system. LoRa backscatter consumes three orders of magnitude lower power than LoRa radios and would significantly extend the battery life. Further it can easily operate for more than 10 years on button cells and printed batteries that are fraction of the size of batteries used with LoRa radios.

• *Cost and size.* Fig. 3(b) plots the cost of active radio and backscatter communication solutions. Radios are at least an order of magnitude more expensive compared to backscatter solutions. This is because an active radio requires analog RF components such as local oscillators, mixers and amplifiers that consume significant silicon area. Since, the cost of an IC is directly proportional to silicon area, the analog components increase the cost of radios. Additionally, radios require external components such as crystals, matching inductors and decoupling capacitors all of which increase the overall cost. In contrast, backscatter solutions are primarily digital in nature and scale with Moore's law that significantly reduce the area and consequently the cost of the backscatter IC to few cents. Also, backscatter communication modules can be built with just an IC, printed antenna and tiny battery and do not require external components like crystals, inductors and decoupling capacitors which reduces the component, manufacturing and assembly costs as well as the overall size.

| Technology | Data Rates | Deployment # 1 ($d_1$, $d_2$) | Deployment # 2 ($d_1$, $d_2$) |
|---|---|---|---|
| Passive Wi-Fi [36] | 1-11 Mbps | 2 m, 30 m | 4.5 m, 4.5 m |
| Interscatter [34] | 2-11 Mbps | 1 m, 27 m | N/A |
| BLE backscatter [30] | 1 Mbps | 0.9 m, 8.5 m | 4.5 m, 4.5 m |
| BackFi [51] | 1-5 Mbps | N/A | 5 m (Full Duplex) |
| LoRea [60] | 2.9 kbps | 1 m, 225 m | N/A |
| HitchHike [68] | 0–300 kbps | 1 m, 50 m | N/A |
| FSK Backscatter [61] | 1.2 kbps | 3 m, 268 m | 25 m, 25 m |
| **LoRa backscatter** | **50 bps–37.5 kbps** | **5 m, 2.8 Km** | **237.5 m, 237.5 m** |

Table 2. **Comparison of LoRa backscatter with existing backscatter systems.** $d_1$ is the distance between the backscatter device and the RF source. $d_2$ is the distance between the backscatter device and the receiver.

• *Sources of Power.* Flexible thin film printed batteries are an emerging source of power which result in cheap, thin and small form factor solutions. However, a key challenge with printed batteries is that they have very limited capacity and peak current requirements. Radios due to their high active power consumption and peak current are incompatible with such technologies. Thin small inexpensive solar cells and button cells, which cost only 10 cents, also suffer from similar limitations. However, LoRa backscatter consume tens of microwatts of power and can operate for ten years on button cells and 10 $cm^2$ area printed batteries. Additionally, LoRa backscatter can be powered with a tiny 2 $cm^2$ size photodiode which makes the overall connectivity module small and inexpensive.

In summary, LoRa backscatter provides the first connectivity solution at a fraction of the cost, size and power consumption of radios while providing reliable and long ranges and being compatible with emerging thin film printed battery solution and inexpensive button cells and tiny solar cells.

*Comparison to prior backscatter research.* LoRa backscatter builds on recent work on backscatter communication [29, 31, 33, 41, 46, 66, 67]. The closest to our work is recent work that backscatters ambient signals like TV [40, 49], Wi-Fi [35, 36, 51, 68, 69], Bluetooth [30, 50, 69] and ZigBee [34, 50]. A comparison of the LoRa backscatter system with existing backscatter systems in terms of data rates and ranges is shown in Table 2. In the table, deployment # 1 refers to the scenario where the backscatter device is placed near the signal source ($d_1$) and the receiver is placed at a distant location ($d_2$). In deployment #2, the backscatter device is equidistant to the signal source and receiver ($d_1 = d_2$) and demonstrates the practical operating range of backscatter systems.

As seen from Table 2, existing backscatter systems have a range of 10 m, when the backscatter device is 6 m away from the RF source and 50 m when the backscatter device is 1 m away [68]. [60, 61] uses FSK modulation to increase the backscatter range. Specifically, the authors show that when the backscatter tag is 3 m from the RF source, the receiver can be 246 m away. In more realistic scenarios, when the tag is around 25 m from the signal source and the receiver, their packet error rate was 100%. In a similar setup, we can deliver packets even when the receiver is 2.8 km away. Our design uses CSS modulation and achieves much higher sensitivities. As a result, we achieve 1–2 orders of magnitude higher communication ranges and thus enable wide-area backscatter.

## 3 SYSTEM DESIGN

LoRa backscatter uses chirp spread spectrum (CSS) to design a wide-area backscatter communication system. In this section, we first provide an overview of CSS. We then present our hybrid analog-digital backscatter design to create CSS transmissions as well as our harmonic cancellation mechanism. We then describe how to synthesize CSS packets that are compatible with the LoRa physical layer by reverse-engineering LoRa. Finally, we outline a link-layer protocol that enables multiple CSS backscatter devices to co-exist with each other.

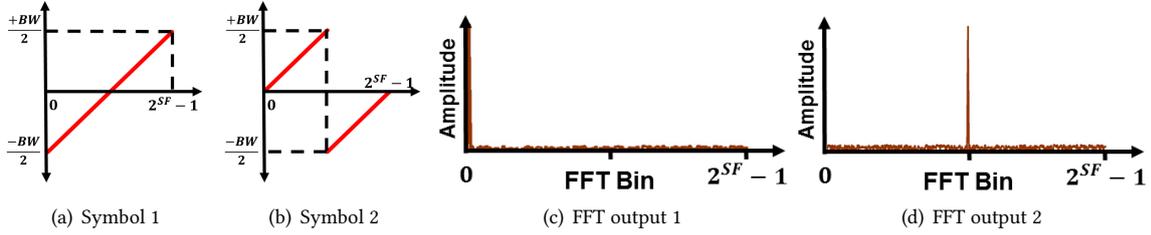

Fig. 4. **CSS modulation & Receiver FFT processing.** The two figures on the left shows two chirp symbols, where the second symbol is a cyclic shifted version of the first, offset by half the chirp duration. $BW$ and $SF$ are the bandwidth and spreading factor respectively. The time delays in the chirp translate to a peak in the FFT domain. The two plots on the right show the FFT output corresponding to the two symbols.

### 3.1 Understanding CSS

**CSS modulation and demodulation.** Chirp spread spectrum (CSS) uses linear frequency modulated chirp pulses to convey information. A key characteristic that CSS leverages is that a time delay in the chirp signal translates to a frequency shift at the output of the FFT. CSS modulation uses this to encode data as cyclic time shifts in the baseline chirp. Figures 4(a) and 4(b) show two chirp symbols, where the first symbol is represented by the baseline chirp and the second is represented by a cyclic shifted chirp, offset by half the chirp duration.

The receiver demodulates these symbols by first multiplying the incoming signal with the baseline chirp and then performing an FFT. Since multiplication in the time domain is correlation in the frequency domain, the resulting operation results in a peak in the FFT frequency bin corresponding to the time delay in the received chirp. Figures 4(c) and 4(d) show the resulting FFT for the two chirps. The figure shows that the FFT has a peak in the first FFT bin for the first symbol (corresponding to zero time delay) and has a peak in the middle FFT bin for the second symbol (corresponding to half a chirp delay). Thus, by tracking FFT peaks, we can decode the data.

Note that one can transmit multiple bits within each chirp symbol. Specifically, say the receiver performs a $N$ point FFT. It can distinguish between $N$ different cyclic shifts which result in a peak in each of the $N$ FFT bins. Thus, we can transmit $log_2 N$ bits within each chirp. In the figure, $N$ is set to $2^{SF}$, where $SF$ is the spreading factor of CSS modulation, which we discuss next.

**CSS parameters and bit rates.** There are three parameters that determine the bit rate achieved while using CSS modulation: 1) chirp bandwidth, 2) spreading factor and 3) symbol rate. As shown in figs. 4(a) and 4(b), if $BW$ denotes the bandwidth, the frequency of the baseline chirp increases linearly between $\frac{-BW}{2}$ and $\frac{+BW}{2}$. The data bits are encoded as cyclic shifts of this baseline chirp, where each cyclic shift represents a modulated symbol. The spreading factor, $SF$, is the number of bits encoded in each chirp duration. From earlier discussion, a chirp with $N$ samples can encode $log_2 N$ bits. Thus, a CSS chirp with a spreading factor $SF$ has $2^{SF}$ samples. Finally, the symbol rate is the number of chirp symbols per second.

With a bandwidth of $BW$, the Nyquist sampling rate is $\frac{1}{BW}$ samples per second. Thus, given a spreading factor of $SF$, the length of each symbol is given by $\frac{2^{SF}}{BW}$ seconds and so the symbol rate is $\frac{BW}{2^{SF}}$ symbols per second. Since each chirp can represent $SF$ bits, the bit rate can be written as, $\frac{BW}{2^{SF}}SF$. Thus, one can achieve different bit rates by either changing the bandwidth or the spreading factor. As we see in §3.4, LoRa also uses error coding codes on top of CSS modulation, giving it a third degree of freedom, in addition to chirp bandwidth and spreading factor, to adapt bit rate.

### 3.2 Synthesizing CSS with Backscatter

**Challenge.** The key challenge is that the complexity of generating CSS signals in the digital domain scales exponentially with the spreading factor used in the CSS transmissions. To understand this, consider CSS modulation with a spreading factor of two. As described earlier, a CSS signal with a spreading factor $SF$ can have $2^{SF}$ cyclic shifts. Thus, the four cyclic shifts shown in Fig. 6 correspond to CSS modulation with a spreading factor of two. To create this signal, the backscatter device needs to generate at least the four frequencies, $f_0, \cdots, f_3$, shown in the figure. More generally, to synthesize a CSS modulation with a spreading factor of $SF$, the backscatter device has to create signals at $2^{SF}$ frequencies. As we see in §3.4, LoRa receivers use spreading factors between 6 and 12. This translated to 64−4096 frequencies. Backscattering all these frequencies requires either using 64−4096 oscillators or running the digital clock at a frequency of $lcm(f_0, f_1, \cdots, f_{4096})$. The first approach is expensive and power consuming while the latter requires using a clock frequency of GHz, which is power consuming and defeats the purpose of using backscatter.

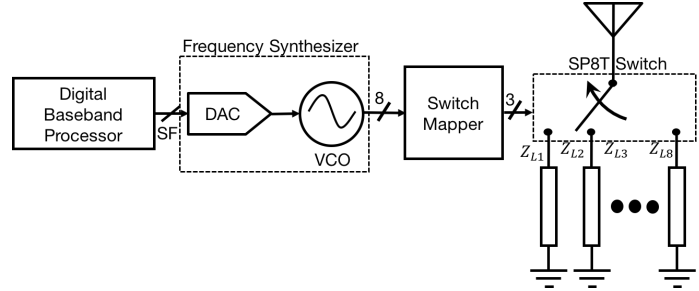

Fig. 5. **Hybrid analog-digital backscatter.** The digital baseband processor generates the frequency plan that is converted in the analog domain to control the output frequency of a VCO. The time shifted versions of the VCO output are mapped according to the dataset of the approximated exponential signal to their respective backscatter impedance values using a SP8T RF switch.

**Our solution.** We present the first backscatter design that can generate CSS modulated signals. Our intuition is to imitate radios [54, 64]. Specifically, we use a hybrid digital-analog backscatter design where we use the energy efficient digital domain to create a frequency plan for the continuously varying CSS signal and then map it to the analog domain using a low-power DAC. For example, to create the second cyclic shift in Fig. 6, the digital baseband creates the frequency plan $f_1, f_2, f_3, f_0$, which the analog domain uses to create the desired frequencies.

Fig. 5 shows the architecture for our backscatter design. It has the digital baseband processor, digital to analog converter (DAC) and a voltage-controlled oscillator. The voltage-controlled oscillator (VCO) is a device that outputs a clock with a frequency that is proportional to the input voltage. We vary the frequency output of the oscillator by using the DAC to generate the appropriate voltages. Specifically, the digital baseband processor outputs an $SF$ bit number, where $SF$ is the CSS spreading factor. This allows us to output $2^{SF}$ voltage levels at the output of the $SF$-bit DAC. The analog voltage output of the DAC controls the frequency of the VCO.

The challenge however, is that, as shown in figs. 4(a) and 4(b), in a CSS encoded packet, the frequency of the signal varies from a negative frequency ($\frac{-BW}{2}$) to a positive frequency ($\frac{BW}{2}$). A voltage-controlled oscillator however only outputs signals at positive frequency. So, we need a mechanism to synthesize negative frequencies using backscatter. From basic communication theory, negative frequencies essentially can be written as complex signals. Specifically, the complex exponent, $e^{j2\pi(\pm f)t}$ can be written as $cos2\pi ft \pm jsin2\pi ft$. Thus, generating negative frequencies requires us to generate both the in-phase cosine signal as well as the out-of-phase sine signal at the desired frequencies. To do this, existing solutions approximate the sine and cosine signals using the square wave output by the VCO. This however results in out-of-band harmonics. In the next section, we describe our harmonic cancellation mechanism.

We note that the receiver receives both the single-tone signal from the RF source as well as the backscattered LoRa packets. Since the backscatter signal is much weaker, the single-tone creates in-band interference. To address this, we shift the single-tone outside the desired band and transform it into out-of-band interference.

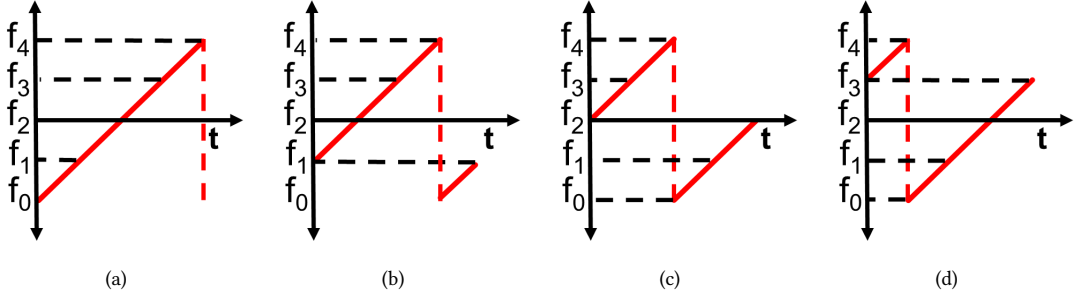

Fig. 6. Four CSS symbols when spreading factor is 2.

At a high level, if the RF source transmits the single-tone signal $e^{2\pi f_c t}$ and the backscatter signal generates the complex signal, $e^{2\pi(\Delta f + f_{LoRa})t}$, then the resulting backscattered RF signal is $e^{2\pi((f_c + \Delta f) + f_{LoRa})t}$. Here $\Delta f$ is a small fixed offset and $f_{LoRa}$ is the varying frequency corresponding to the baseband LoRa modulation. The above equation shows that by using a small frequency offset, $\Delta f$, we can create the LoRa signals in a band centered at $f_c + \Delta f$ which is different from that of the single tone, $f_c$.

### 3.3 Backscatter Harmonic Cancellation

As described earlier, prior backscatter designs [34, 36] use square waves to approximate sine and cosine waves which results in harmonics. To understand why this happens, we recall that a square wave at a rate of $\Delta f$ can be written as a sum of cosine waves using the following expression:

$$Square(\Delta f t) = \frac{4}{\pi} \sum_{n=0}^{\infty} \frac{1}{2n+1} cos(2\pi (2n+1) \Delta f t)$$

If the RF source transmits $cos(2\pi f_c t)$ and the backscatter device is switching with a square wave operating at $\Delta f$ frequency, signal transmitted by the backscatter device can be written as $cos(2\pi f_c t)Square(\Delta f)$. As a result, in addition to generating the desired signal at $f_c + \Delta f$, the above operation also generates the mirror copy at $f_c - \Delta f$, 9.5 dB lower harmonic at $f_c \pm 3\Delta f$, 15 dB lower harmonic at $f_c \pm 5\Delta f$ and additional lower power harmonics. Recent work [34] has demonstrated how one can eliminate the mirror copy being generated at $f_c - \Delta f$ using single side band backscatter technique. However, this technique still preserves the third, fifth and other odd order harmonics. The third and fifth order harmonics are only 9.5 and 15 dB lower than the desired backscattered signal and hence create interference on the wireless channel. More importantly, since the LoRa protocol has very low sensitivities, LoRa devices operating in channels overlapping with the third and fifth harmonics experience in-band interference from backscatter devices.

***Our Solution.*** Our insight is to use a different signal from the square wave to approximate a cosine and sine wave. On a high level, one can think of an analog signal as a discrete signal with infinite distinct voltage levels and smooth transitions, which results in a clean spectrum without any harmonics. However, square wave has only two levels with discontinuous step transitions, which results in high frequency components. Our key idea with harmonic cancellation is to introduce additional voltage levels to better approximate a sinusoidal signal, by imitating radios [53, 63], and obtain a cleaner frequency spectrum.

Consider the approximation of a cosine wave in Fig. 7, using a signal with four voltage levels. This approximated cosine wave can be written as the sum of three square waves slightly shifted from one other, $S_0(t)$, $S_1(t)$ and $S_2(t)$, as shown in the figure. Here $T$ is the time period.

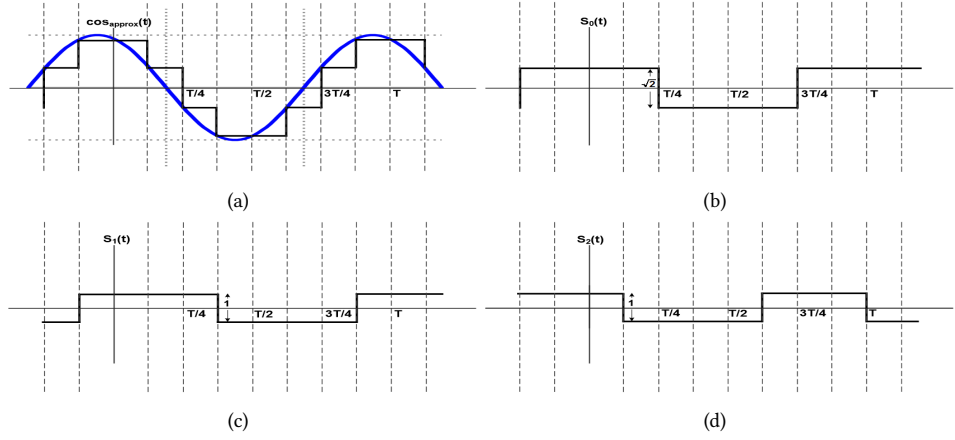

Fig. 7. **Approximation of a cosine wave with multi-level signal.** We approximate the cosine wave as a sum of three digital signals $S_0(t)$, $S_1(t)$ and $S_2(t)$ resulting in a multi-level signal.

$$S_0(t) = \frac{4\sqrt{2}}{\pi} \sum_{n=0}^{\infty} \frac{\sin[(2n+1)2\pi\Delta f(t+\frac{T}{4})]}{2n+1}$$

$$S_1(t) = \frac{4}{\pi} \sum_{n=0}^{\infty} \frac{\sin[(2n+1)2\pi\Delta f(t+\frac{T}{8})]}{2n+1}$$

$$S_2(t) = \frac{4}{\pi} \sum_{n=0}^{\infty} \frac{\sin[(2n+1)2\pi\Delta f(t+\frac{3T}{8})]}{2n+1}$$

Using the above expression for the three signals, we can now express the approximated cosine waveform as,

$$cos_{approx}(2\pi\Delta f t) = S_0(t) + S_1(t) + S_2(t)$$

$$= \frac{4}{\pi} \sum_{n=0}^{\infty} \frac{\sin[(2n+1)2\pi\Delta f(t+\frac{T}{4})][2\cos((2n+1)\frac{\pi}{4}) + \sqrt{2}]}{2n+1}$$

The sine wave can now be generated by simply shifting the cosine waveform by quarter of the time period. Using these approximations for the sine and cosine parts of the waveform, the exponential, $e^{j2\pi\Delta f t}$, can now be mathematically written as,

$$e^{j2\pi\Delta f t} = cos(2\pi\Delta f t) + sin(2\pi\Delta f t)$$

$$\approx cos_{approx}(t) + sin_{approx}(t)$$

$$= \frac{2}{\pi} \sum_{n=0}^{\infty} \frac{1}{2n+1}[2\cos((2n+1)\frac{\pi}{4}) + \sqrt{2}]$$

$$[e^{j(2n+1)2\pi\Delta f t}((-1)^n + 1) + e^{-j(2n+1)2\pi\Delta f t}((-1)^n - 1)]$$

Let us now consider what happens with the above equation for different values of $n$. When $n$ is zero, the term corresponding to the negative frequency in the second parenthesis computes to zero and only the positive frequency is preserved resulting in single side band generation.

$n = 1$ and 2 correspond to the third and fifth harmonic respectively. For these cases $\cos[(2n + 1)\frac{\pi}{4}] = \frac{-\sqrt{2}}{2}$ and so the above equation computes to zero cancelling the third and fifth harmonics. Thus, by using the above four level signal, we can cancel the third and fifth harmonics as well as achieve single sideband modulation at the same time. More generally, when $n$ is of the form $(8k + 3)$ and $(8k + 5)$, $\cos[(2n + 1)\frac{\pi}{4}] = \frac{-\sqrt{2}}{2}$ and hence all the corresponding harmonics will be cancelled. In summary the four-level approximated exponential signal cancels at least the third and fifth order harmonics.

If required, subsequent harmonics can be cancelled by adding more levels. Specifically, addition of each level cancels the next higher order harmonic. For example, five voltage levels cancel the seventh harmonic and ninth harmonic is cancelled with six voltage levels. Finally, every backscatter switch has a finite delay, which provided additional filtering and automatically suppresses higher order harmonics (greater than 9), without the need for additional levels.

In our implementation, we use four levels to cancel the third and fifth order harmonics. We generate the approximated signals on the backscatter device in the digital domain. The four level cosine signal takes one of four values {0.9239, 0.3827, −0.3827, −0.9239}. Fixing the cosine value, lets the sine take one of two values. Thus, the exponential $e^{j2\pi\Delta f t}$ can take one of eight complex values. We create these complex values by leveraging existing backscatter techniques [34, 59] that change the impedance connected to the antenna. Fig. 5 shows the architecture of our system where we switch the antenna between eight different complex impedance values to generate the eight complex values corresponding to our exponential signal. Specifically, we implement the exponential wave approximation in a digital logic block called the switch mapper. It takes as input the eight phases of the VCO output to correspond to the time instances when waveform $S_0$, $S_1$ and $S_2$ and their time shifted versions (corresponding to the sine wave) undergo transitions. We generate the eight phases of the clock signal by just shifting the signal in one-eighth of the time period increments and the switch mapper outputs the 3 control bits which toggle the switch between 8 impedance values to generate the approximate exponential wave. Using this technique we can successfully cancel the mirror image as well as the third and fifth harmonics, thereby improving the spectral efficiency of backscatter systems.

### 3.4 Synthesizing LoRa Packets

LoRa achieves its high sensitivity numbers using CSS modulation. The physical layer specification for LoRa however is proprietary and is not publicly available. So, we reverse-engineer the LoRa physical layer using the patents filed by Semtech [26, 56], which is the key LoRa chipset manufacturer. We also use the Semtech 1276 starter kit [20] that provides an interface to transmit LoRa packets with various bitrates and an arbitrary payload. Finally, we analyze the transmissions from the LoRa chipsets on a USRP.

*Packet Structure.* Fig. 8 shows the structure of a LoRa packet, in the form of a spectrogram. The figure shows a sequence of repeating chirps at the beginning to represent the preamble. LoRa supports a variable length preamble between 6 and 65535 chirp symbols. To convey the end of the preamble to the receiver, the preamble ends with synchronization symbols and two and a quarter down-chirp symbols where the chirp goes from the positive to negative frequency. After down-chirps, the packet has an optional header with information about the bit rate used. This is followed by a CSS-encoded payload. An optional 16-bit CRC is send at the end of the packet.

*Bit Rates.* LoRa bit rates depend on three parameters: the error correction coding rate, chirp bandwidth and spreading factor. LoRa supports four different hamming code rates and eight chirp bandwidths of 7.8 kHz, 10.4 kHz, 20.8 kHz, 31.25 kHz, 62.5 kHz, 125 kHz, 250 kHz and 500 kHz. Further, the spreading factor can be set independently to one of seven values: 6, 7, 8, 9, 10, 11 and 12. The LoRa hardware allows these three parameters to be independently modified resulting in a total of 224 bit rate settings between 11 bps and 37.5 kbps.

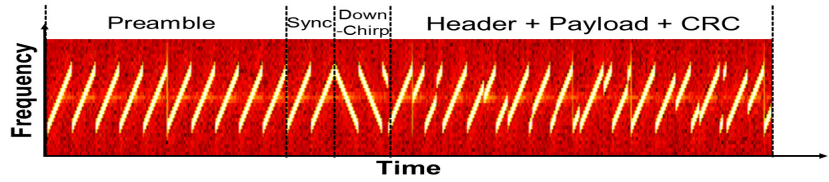

**Fig. 8. LoRa packet structure.**

We use the above packet format to synthesize LoRa packets with backscatter. We note the following.

• LoRa bandwidth and spreading factor are set a-priori and assumed known at the transmitter and receiver. The header can include information about the bit rate and payload size used, but is optional and its overhead can be reduced by statically configuring these parameters, which we do in our backscatter system.

• To achieve a high sensitivity, the phase of the LoRa chirps has to change continuously with time and has the same value at the beginning and the end of the chirp [26, 56]. To achieve this with backscatter, at each frequency, we increase the phase by $\frac{2\pi}{SF}$. This ensures that the phase at the beginning and end of each chirp symbol is the same and so we can maintain phase continuity across chirp symbols.

• To comply with FCC regulations, LoRa uses frequency hopping while using lower-data rate transmissions that occupy significant amounts of time on the channel. Specifically, while using a chirp bandwidth of 125 kHz, LoRa divides the 900 MHz ISM band into 64 channels starting at 902.3 MHz, with increments of 200 kHz. Similarly, with a chirp bandwidth of 500 kHz, LoRa divides the band into 8 channels in increments of 1.6 MHz. The transmitter performs frequency hopping between these channels to transmit data to be compliant with FCC. For backscatter, FCC only regulates the signal source and not the backscatter device [19][2]. Thus, we instead hop the frequency of the single-tone transmitter. The backscatter device however uses the same frequency offset, $\Delta f$ and is oblivious to this frequency hopping mechanism. To ensure that the backscattered LoRa transmissions always lie in the LoRa channels, we hop the frequency of the single-tone source at a constant frequency offset of $\Delta f$ from the LoRa channels. Specifically, to generate the backscatter signals at the LoRa channels, $f_1, f_2, \cdots, f_n$, the single-tone source performs frequency hopping across $f_1 - \Delta f, f_2 - \Delta f, \cdots, f_n - \Delta f$.

### 3.5 Link-Layer Protocol

We describe how the RF source arbitrates the channel between backscatter devices. Then we explore concurrent transmissions from multiple backscatter devices.

*Arbitrating the channel between backscatter devices.* At a high level, we use TDMA to allocate the wireless channel between different backscatter devices. Specifically, the RF source divides time into slots and transmits the single tone signal once in each slot. Each backscatter device only transmits during its assigned slot.

The above protocol requires the backscatter devices to detect the beginning of the single tone from the RF source. To do this, our design uses existing energy detector hardware circuits that consume between 98 nW and 2.4 μW and can detect input signals as low as -71 dBm [32, 52]. Note that this is larger than the -148 dBm LoRa sensitivity. This is however the power of the RF source at the backscatter device, while the latter is the power of the backscattered signal at the LoRa receiver. In all our experiments, the signal strength at the backscatter device was at least -45 dBm. We note that to improve the accuracy of detecting the signal from the RF source, we can also use a preamble signal like an alternating ON-OFF keying sequence of energy and no energy. This reduces the probability of confusing random transmissions in the 900 MHz band for our RF source.

---

[2]As a result, RFID tags do not have an FCC ID, while RFID readers have to get approved by FCC and have an FCC ID.

To synchronize the slots across different backscatter devices, the RF device uses an unique ON-OFF keying sync pattern at the beginning of the TDMA round robin. This allows devices to determine the slot boundaries for a whole round-robin duration. We note that the backscatter devices do not need to have their receivers ON all the time. In particular, depending on the application, the backscatter device only transmits when it has new data. Similarly, the RF source does not transmit the single-tone signal during a time slot, if the corresponding backscatter device is not scheduled.

*Concurrent LoRa transmissions using backscatter.* So far we assume that only a single backscatter device can transmit at a time. However, LoRa divides the 900 MHz band into 64 125 kHz LoRa channels each of which can have a LoRa transmission. Thus, using a single RF source, we can enable up to 64 backscatter devices to transmit concurrently on different LoRa channels to their corresponding receivers. Taking it a step further, CSS transmissions that use different spreading factors on the same LoRa channel are uncorrelated with each other [20]. Thus, in principle, we can have multiple backscatter devices with different spreading factors use the same LoRa band and transmit concurrently to their receivers, increasing the overall network throughput.

We note that we can use the energy detector circuit to transmit ACKs on the downlink channel. However, given the resilience of CSS modulation and the high sensitivity values, performing application level error correction codes is sufficient to deliver data to the receivers, with a high probability, without the need for explicit ACKs.

## 4 HARDWARE IMPLEMENTATION

We first built a proof of concept prototype using commercial off-the shelf (COTS) components and used it to characterize the performance and operating range of our system. Then, we design an integrated circuit based on the hybrid analog-digital architecture for harmonic cancellation proposed in §3.2 to quantify its power consumption.

**COTS implementation.** Our COTS implementation consists of an RF and a baseband section. The RF section is implemented on a four layer FR4 substrate and consists of three cascaded ADG904 [16] switches to create a SP8T switch network. The switch toggles a 2 dBi whip antenna [1] across the eight impedance states required for the harmonic cancellation technique. In our optimized implementation, we use 47 $pF$, 3.3 $nH$||82 $\Omega$, 21 $nH$||680 $\Omega$, 8.2 $nH$||330 $\Omega$, 1.8 $k$ $\Omega$, 1.5 $pF$||56 $k$ $\Omega$, 9.1 $pF$||560 $\Omega$ and 3.9 $pF$ as our impedance values to achieve the desired complex values while incurring a loss of only 4 dB in our backscatter switch network.

We implement the baseband section using the DE0-CV development board for an Altera Cyclone V FPGA [3, 6]. We generate CSS modulated packets in digital domain using Verilog HDL and output square waves corresponding to the real and imaginary components of the signal to the RF section by interfacing the two digital I/Os on the FPGA to the SP8T switch network through level shifters.

**IC implementation.** We design LoRa backscatter in 65 nm LP CMOS process by TSMC [15]. Our IC is composed of three main components: digital baseband processor, frequency synthesizer and the backscatter switch network.

*Baseband Processor.* It takes payload data and packet specifications such as spreading factor, bandwidth and code rate as input and synthesizes the LoRa packet in accordance to the structure described in §3.4. Next, it maps the bits in the packet to a frequency plan that is used by the frequency synthesizer block to create the chirp spread spectrum signal. We describe the behavioral model for LoRa packet in Verilog and use Design Compiler by Synopsis [11] to synthesize the transistor level implementation. Our baseband processor consumes 1.25 $\mu W$ to generate a LoRa packet with spreading factor of 12, 31.25 kHz bandwidth and (8,4) hamming code.

*Frequency Synthesizer.* We integrate the DAC and VCO in Fig. 5 into a single frequency synthesizer block. A frequency synthesizer or phase locked loop (PLL) takes a low frequency clock source as input and up converts it to a higher frequency. The ratio of the output frequency to the reference frequency is set by a divide ratio. We use the baseband processor output to directly control the divide ratio of the frequency synthesizer and modulate the output frequency to generate CSS modulated data. We use Johnson counter to generate the four versions of the clock which are shifted by one eighth of the time period. The four shifted versions and their complements are used to represent the eight phases of the clock signal. The frequency synthesizer consumes 4.5 $\mu W$ to generate 31.25 kHz CSS packets with a spreading factor of 12 and a 3 MHz offset.

*Backscatter Switch Network.* The backscatter switch network takes the eight phases of the clock (four time shifted versions and their complements) and maps the eight phases to RF switches corresponding to their respective impedance values. The switches are implemented using NMOS transistors that toggle the antenna between eight discrete impedance states consisting of resistors and capacitors. We limit the impedances to only resistors and capacitors since inductors consume huge area and are prohibitively expensive in IC's. This results in a more constraint constellation map and 3 dB loss in backscattered signal but is a reasonable compromise for low cost. During active operation, the backscatter switch network consumes 3.5 $\mu W$ to backscatter CSS modulated packets at 3 MHz offset. In total, the IC consumes 9.25 $\mu W$.

## 4.1 Cost Analysis

Backscatter communication systems consume 3–4 orders of magnitude lower power than their counterpart radios [2, 30, 34, 36, 55, 59]. This is because instead of using complex and power hungry analog front end components of radios, backscatter systems use switches to modulate reflections. As a result, backscatter systems are not only extremely low power, but also consume significantly smaller area. This is a key benefit of backscatter systems because for a given technology node, the cost is directly proportional to the die area. Since active radios (e.g. Wi-Fi, LoRa, BLE) are composed of complex and area intensive components such as power amplifiers, mixers, local oscillators operating at RF frequencies, the RF front end in these systems (excluding the digital baseband) typically occupies about 10 $mm^2$ of die area [22, 28, 37, 38, 45]. In contrast, backscatter systems such as RFID tags are much simpler and occupy significantly lower area. The typical die area for the communication module in RFID based backscatter systems is in the order of 0.15 $mm^2$ [29, 44, 62, 66, 67]. As a result, the cost of a backscatter chip is at least 60 times lower.

Our LoRa backscatter system uses a similar architecture to an RFID tag with some key additions. We use an All-Digital Phased-Lock Loop (ADPLL) and an eight-stage backscatter network implemented using area efficient resistor and capacitors (metal in metal capacitors are implemented on higher metal layers on top of existing semiconductor structures) which occupies less than 0.01 $mm^2$. Therefore, the die area and the cost structure for our communication module would be similar to that of an RFID tag and would be at least 60 times less than active radios.

## 5 EVALUATION

We first characterize the LoRa receiver at different bit rates and sensitivities and evaluate its ability to decode the backscattered signals in the presence of single tone interference from the RF source. Then we evaluate the operating range as well as our harmonic cancellation technique.

## 5.1 Receiver Characterization

As described in §3.1, the data rate and sensitivity of a CSS modulated packet is determined by the spreading factor, bandwidth and coding rate. In this section we will use the Semtech SX1276 receiver and evaluate the trade-off between sensitivity and data rates supported by the hardware.

To do this, we use a wired experimental setup to ensure that variations due to multipath do not impact our results. We use two SX1276 chipsets and configure them to be both the CSS transmitter and receiver. We connect the antenna port on the two chipsets using variable attenuators and an RF cable. As described in §3.4, LoRa supports 224 bit rate configurations between 11 bps and 37.5 kbps. We implement seven rates spread across

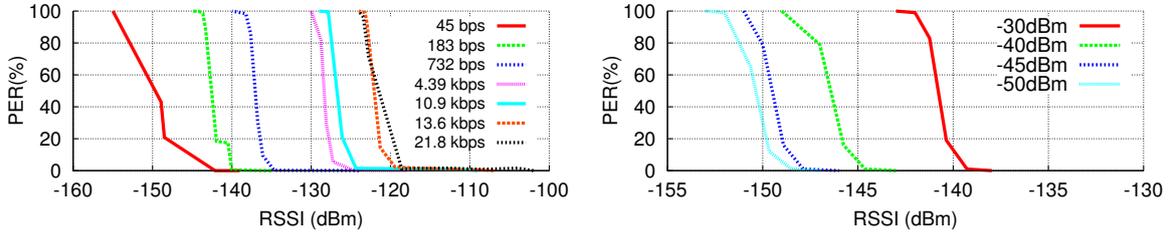

(a) **Receiver Characterization.** PER as a function of RSSI values reported by the receiver for different data rates.

(b) **Resilience to Out-of-band interference.** PER as a function of RSSI values for different interference power levels.

Fig. 9. We characterize our receiver by measuring PER as a function of reported RSSI in presence and absence of out of band interference.

21.8 kbps and 45 bps by picking the appropriate spreading factor, bandwidth and coding rate. We use variable attenuators to change the power of the received packet. Specifically, for each of the data rates, we start with a low attenuation value where we receive all packets and increase the attenuation till we stop receiving all packets. We transmit 1000 packets with eight byte payload and two byte CRC and measure the packet error rate (PER).

Fig. 9(a) plots the PER as a function of the RSSI values reported by the receiver chipset for the seven data rates. As expected, we observe that the sensitivity is inversely proportional to the data rate of the packet. For the highest data rate of 21.8 kbps, receiver can decode packets with PER of less than 1% up to an RSSI value of -119dBm while for the lowest tested data rate of 45 bps we receive packets down to an RSSI of -142 dBm. Below an RSSI value of -142 dBm, the receiver cannot correctly decode packets for any of the tested data rates. Finally, we note that similar to Wi-Fi, the RSSI values reported by the chipset do not correspond to the actual power level of the received packet. However, the reported RSSI value is a good indicator and is proportional (though not linear) to the power level of the received packet.

## 5.2 Resilience to Out-of-band interference

Next, we evaluate how well the receiver can decode backscattered packets in the presence of out-of-band interference from the RF source. Specifically, the RF source transmits at a frequency offset from the backscatter signal and hence, can create out-of-band interference. To check this, we setup the following test bench: we connect the RF source, backscatter prototype and the receiver using a circulator setup to isolate the results from multipath. Specifically, we use a variable attenuator and connect the RF source transmitting the single tone at 905 MHz to the first port of the circulator. The LoRa backscatter hardware prototype is connected to the second port using another attenuator. The SX1276 receiver configured to receive packets at 906 MHz is connected to the third port of the circulator with an RF cable. We set the LoRa backscatter device to backscatter packets with a spreading factor of 12, bandwidth of 31.25 kHz and a (8,4) hamming code with 8 byte payload and 2 byte of CRC at a frequency offset of 1 MHz.

The out of band interference experienced by the receiver is a function of the distance between the RF source and the receiver and depends on the deployment scenario. To cover different scenarios, we use the attenuator corresponding to the RF source to vary the power of the interference experienced by the receiver. We set the power of the interference at the receiver to -30 dBm, -40 dBm, -45 dBm and -50 dBm which translates to distances of 50 m, 200 m, 300 m and 500 m respectively between the RF source and receiver in free space. Next, we change the attenuator corresponding to the backscatter device to decrease the power of the backscattered packet. We measure PER as a function of the RSSI values reported by the receiver for different interference power level. Fig. 9(b) plots the results which show the following:

• The receiver sensitivity is inversely proportional to out of band interference. When the interference is -30 dBm, the receiver can decode packets down to an RSSI value of -139 dBm. As this interference reduces to -45 dBm and

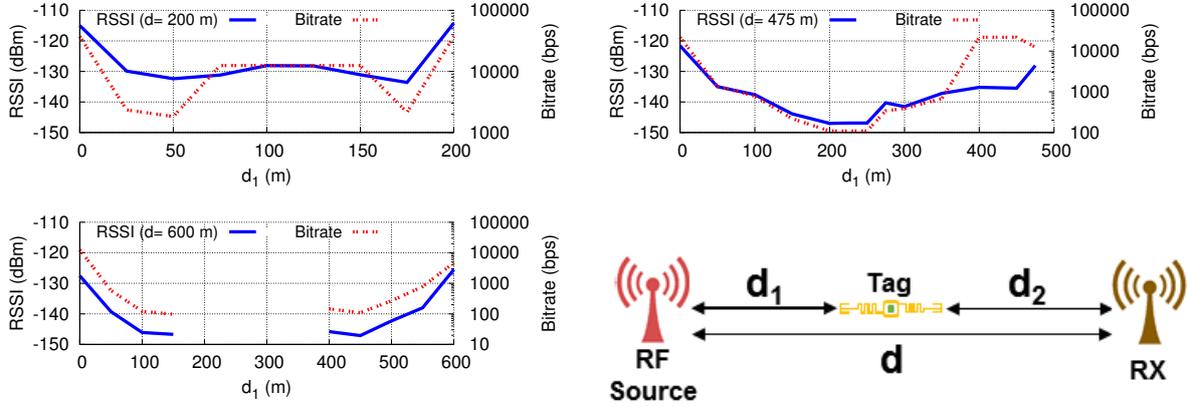

Fig. 10. **RSSI in Deployment scenario 1.** $d$ is the distance between the RF source and receiver. We move the backscatter device along the line joining them. This figure also shows the corresponding LoRa bitrate at which we successfully receive all our ten packets from the backscatter device without any loss.

-50 dBm, which is closer to values seen in our deployments, the receiver can correctly decode packets down to RSSI values of -146 and -148 dBm respectively.

• With out-of-band interference, the receiver can decode packets at lower RSSI numbers than in §5.1. This is because the RSSI values reported by the chipsets are just an approximate indicator of link quality and do not represent the absolute measure of power or sensitivity.

• In setup with -40dBm and -50 dBm out of band interference, an offset of 1 MHz at the backscatter device achieves sensitivities that were within a few dB of the theoretical limit. We can increase the offset for better sensitivities when facing a higher interference power.

## 5.3 Operational Range

We evaluate the operating range of the LoRa backscatter system in two different scenarios. We use the SX1276 LoRa development kit connected to a power amplifier as the RF source and configure it to transmit 30 dBm into a 6 dBi patch antenna at 915 MHz. This is the maximum power permitted by FCC on the 900 MHz ISM band. We configure the backscatter device to transmit LoRa packets at a frequency offset of 3 MHz with a spreading factor of 12, bandwidth of 31.25 kHz and a (8,4) hamming code with 3-byte payload and 2 byte CRC. The SX1276 transceiver chip decodes packets received at 918 MHz and we log the RSSI values for packets that pass CRC.

*Deployment scenario 1.* The first scenario we consider is a deployment where the RF source and the receiver are separated from each other and the backscatter device can be at any location between them. To test this, we place the RF source and the receiver at a distance $d$, as shown in Fig. 10, and move the backscatter device in a straight line between them. At each location of the backscatter device, we measure the RSSI value reported by the receiver. Due to the large operating range, we ran the experiments on a straight road next to open fields. Fig. 10 plots the RSSI values for three different distances between the signal source and the receiver. The x-axis represents $d_1$, the distance between the RF source and the backscatter device. The plots show the following,

• We get a low RSSI when the backscatter device is at the midpoint between the RF source and the receiver. This is because the signal from the RF source attenuate as $1/d_1^2$ before arriving at the backscatter device. The signals generated by backscattering these attenuated signals further attenuate as $1/d_2^2$. Thus, the backscatter signal strength at the receiver scales as $1/d_1^2 d_2^2$ which is minimum when $d_1 = d_2$.

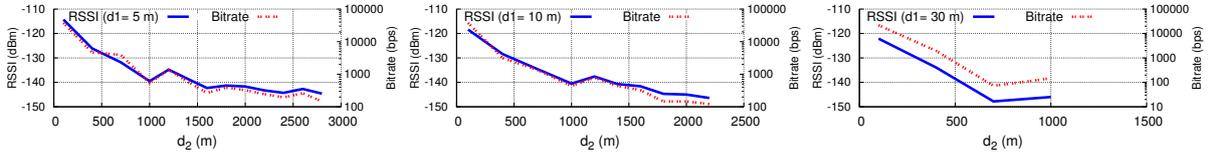

Fig. 11. **RSSI in Deployment scenario 2.** $d_1$ ($d_2$) is the distance between the backscatter device and RF source (receiver). We fix the location of the backscatter device and RF source and move the receiver away from the backscatter device.

• Our system operates at all locations up to a maximum separation of 475 m between the RF source and the receiver. At 600 m, the backscatter device only works close to either the RF source or the receiver. This shows that our design has an operational range of 475 m.

*Deployment scenario 2.* In the second scenario, we fix the distance between the backscatter device and the RF source and move the receiver away from the backscatter device and measure RSSI. This is a scenario where the backscatter device is close to the RF source — an example deployment is one where the sensor is in the wall and the RF source is placed nearby where it can be plugged in. Our initial experiments in this setup showed that we could achieve very large ranges. So we test this scenario on the same road, which spans multiple kilometers. Specifically, we set the RF source and backscatter device on the roadside. We change the separation between the RF source and the backscatter devices between three values of 5, 10 and 30 meters. For each separation value, we drive the receiver away from the backscatter device, alongside the road and measure the RSSI of CRC passed packets at the granularity of 200 m.

Fig. 11 plots RSSI as a function of the distance between the backscatter device and the receiver. It shows results for three different distances between the RF source and the backscatter device. The plots show that the receiver can receive packets at a distance of 2.8 km from the backscatter device, when the backscatter device and the RF source are separated by 5 m. We note that the receiver could decode all transmitted packet at all reported locations. When we increase distance between the RF source and the backscatter device, the operating distance reduces. At 30 m separation, the receiver could receive packets up to 1 km.

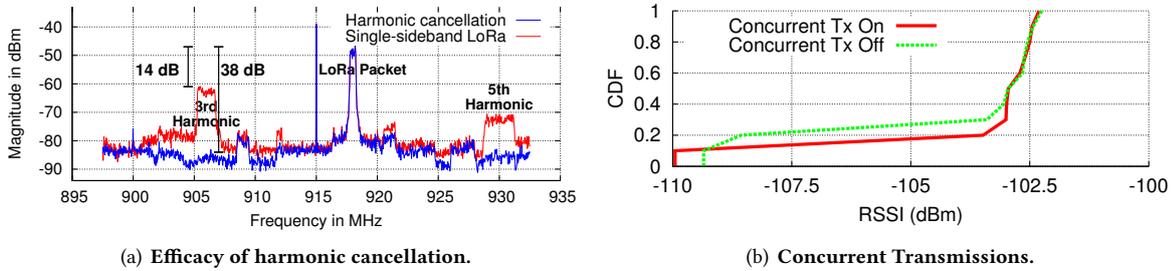

(a) **Efficacy of harmonic cancellation.**

(b) **Concurrent Transmissions.**

Fig. 12. The plot on the left demonstrates the efficacy of our harmonic cancellation technique. The right plots CDF of RSSI of a LoRa backscatter device in the presence and absence of another LoRa backscatter device concurrently transmitting in adjacent band.

## 5.4 Efficacy of harmonic cancellation

Next, we evaluate how well our harmonic cancellation technique works in practice. To do this, we capture the spectrum of the signal generated using our harmonic cancellation backscatter hardware and compare it to a baseline hardware that does not implement harmonic cancellation. Specifically, for the baseline we replicate the state-of-the-art single-sideband Wi-Fi hardware from [34] and adapt it using the techniques in §3.2 to generate LoRa packets. We use the Semtech SX1276 development kit to generate the single tone signal at 915 MHz. Since the third and fifth harmonics span a wide band, we use a spectrum analyzer to capture the signal. To avoid interference from other wireless devices, we connect the Semtech kit to port 1 of a circulator, the backscatter hardware to port 2 and measure the spectrum of the backscattered signal by connecting a spectrum analyzer to the third port of the circulator. We set our backscatter hardware to transmit LoRa packets with spreading factor of 6, bandwidth of 500 kHz and a (8,4) hamming code with 8 byte payload and 2 byte CRC at a 3 MHz offset.

The red line in Fig. 12(a) shows the spectrum of the backscattered LoRa packet with the single-sideband hardware. The plot shows a single tone signal at 915 MHz, which is the RF source and the desired LoRa packet at 918 MHz. In addition, we see the third and fifth harmonics at 906 MHz and 930 MHz respectively. These are 14 dB and 22 dB lower than the desired backscatter LoRa packet. These numbers are 4.7 dB and 7 dB lower than our analysis with square wave in §3.2 because in frequency modulated systems, the higher order harmonics are spread across frequency proportional to the order of the harmonic. Note that the third and fifth harmonics occur on opposite sides of the desired LoRa transmission at 918 MHz because of single-sideband architecture. The key observation however is that the third and fifth harmonics create interference in the 900 MHz band.

The blue line shows the spectrum of the backscattered LoRa packet with our harmonic cancellation backscatter hardware. The spectrum shows that the third (906 MHz) and fifth harmonics (930 MHz) are 38 dB lower than the backscattered LoRa packet at 918 MHz and is close to noise. This demonstrates the efficacy of our harmonic cancellation technique. We note that while in theory we should get perfect cancellation, in practice backscatter hardware built using switches and passive components have tolerances and variances that introduce small errors.

## 5.5 Concurrent Backscatter Transmissions

Finally, we present a proof-of-concept evaluation of LoRa backscatter when multiple backscatter devices concurrently transmitting signal from a single RF source. We use two backscatter devices and configure them to continuously transmit packets at different frequency offsets of 0.75 MHz and 1 MHz. Since the two devices create transmissions on different LoRa bands, they can concurrently transmit to their receivers. We set the RF source to transmit a single tone at 905 MHz and deploy the system in a large atrium measuring 104 by 32 feet. We place the RF source and the two receivers at either ends of the room. We fix the location of the first backscatter device at the center of the atrium. We move the second backscatter device (operating at 1 MHz offset) across ten different locations of the atrium. Fig. 12(b) shows the CDF of the RSSI values of packets transmitted by the first backscatter device in the presence and absence of concurrent transmissions by the second device. The plots show that concurrent transmissions have negligible impact on the performance of the first backscatter device.

## 6 APPLICATION DEPLOYMENTS

We show two classes of applications enabled by LoRa backscatter. The first set of applications home/office sensing and precision agriculture leverage the large operational range of our communication system. We then demonstrate the use of LoRa backscatter in applications with unfavorable RF propagation conditions such as implantable and epidermal devices. In all our application deployments, we set the backscatter device to transmit with a spreading factor of 12, bandwidth of 31.25 kHz and a (8,4) hamming code with three byte payload and two byte of CRC. We use the SX1276 receiver to log the RSSI value of packets that pass the CRC.

### 6.1 Wide-Area Applications

We deploy our system in two wide-area applications: home sensing and precision agriculture and evaluated the system in real world deployment scenarios.

*6.1.1 Whole-Home and Office Sensing.* Prior backscatter techniques such as [35, 36] have tried to replace power-consuming radios in home sensing applications with backscatter but suffer from limited range and do not provide the coverage required for a practical home deployment. We deploy our system in a 4,800 $ft^2$ (446 $m^2$) home spread across three floors in a major metropolitan US city. The layout and floor plan of the house is shown in Fig. 14. We put the RF source at one corner of the house on the third floor and place the receiver in the basement. We move the backscatter device across the three floors of the occupied house in 6 x 6 foot grid increments and report RSSI of the received packets at each location.

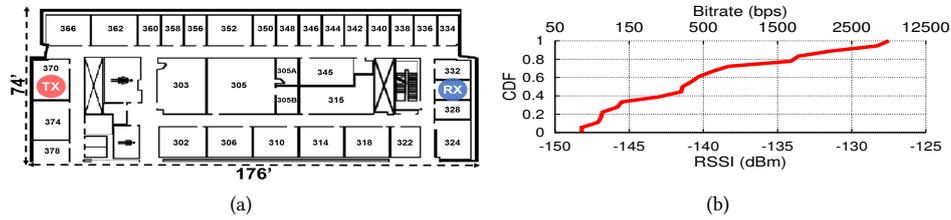

(a)                                                          (b)

Fig. 13. **Office Deployment.** We receive backscattered packets across one floor of an office building with 13,024 $ft^2$ (1210 $m^2$) area.

Fig. 14 plots the RSSI of the received backscattered packets at each of the three floors as well as the outside lawn area. We use a 1 MHz offset on the backscatter device. Our results show that the across the entire house, our system was able to achieve RSSI values greater than -144 dBm which translates to reliable wireless coverage across the entire house at 45 bps, with a single RF source and receiver. These rates are sufficient for most home automation devices such as temperature sensors that transmit small packets sporadically.

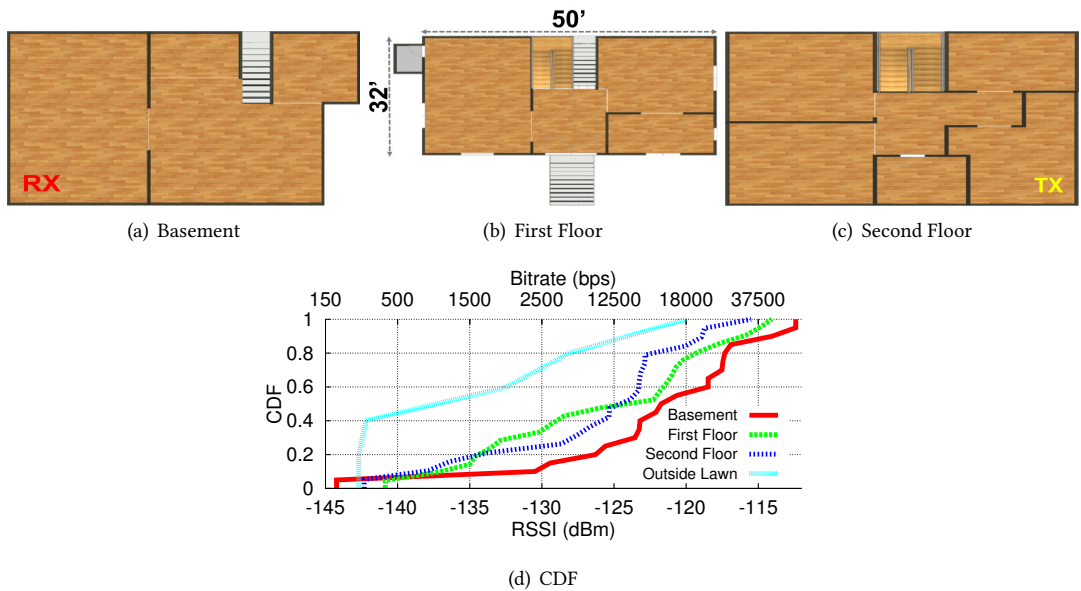

(a) Basement          (b) First Floor          (c) Second Floor

(d) CDF

Fig. 14. **Home Deployment.** We receive packets across the 4,800 $ft^2$ (446 $m^2$) house spread across three floors.

Next, we deploy our system with a 3 MHz offset on the backscatter devices in an office space spanning 13024 $ft^2$ (1210 $m^2$). As shown in Fig. 13, the deployment spans 41 offices that are separated by double sheet-rock (plus insulation) walls with a thickness of approximately 5.7 inch (14.5 $cm$). We place a RF source and receiver on the two end of the space as shown in the figure. We note that the receiver is behind a heavily insulated set of rooms including the restroom and supply closets with concrete and metal structures separating them. We move the backscatter device into different offices across the whole floor in 26 by 26 foot grid increments and report the RSSI of received packets at each location. Fig. 13 plots the CDF of the RSSI of received backscattered packets across the whole space, demonstrating wide-area backscatter coverage in significant multi-path environments.

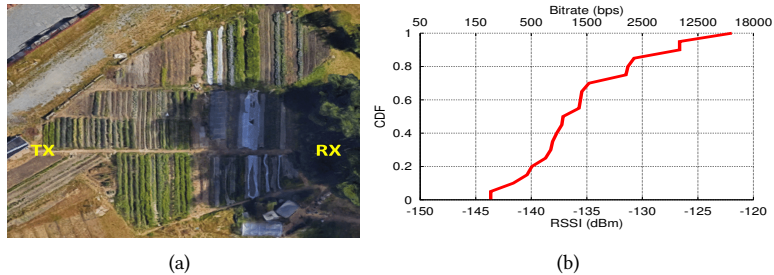

(a)                    (b)

Fig. 15. **Precision Agriculture.** We receive backscattered packets across a one-acre ($4046\ m^2$) farm.

*6.1.2 Precision Agriculture.* We deploy LoRa backscatter in a one-acre ($4046\ m^2$) vegetable farm owned by our organization. We set the RF source and the receiver at the opposite ends of the farm, whose aerial view is shown in Fig. 15. We divide the farm into 45 ft by 45 ft grids and placed the backscatter device with a 1 MHz offset at ground level between plants and bushes. We measure RSSI values by placing the backscatter device across 20 locations at the center of the grids. These locations are a combination of line of sight and non-line of sight positions to the RF source and/or receiver due to presence of green houses and storage sheds across the farm.

Fig. 15 plots the CDF of the RSSI of the packets received across the entire farm. The plot shows that the median RSSI is -137 dBm and the minimum RSSI was -143 dBm. This demonstrates that using a single RF signal source and receiver, we could provide reliable backscatter communication across a one-acre ($4046\ m^2$) farm. We note that typical farms are much larger and might require more RF signal sources and receivers. However, since Semtech SX1276 transceiver is relatively cheap and the cost of the backscatter devices is in the order of a few cents, we could achieve cost and power saving by deploying backscatter in agriculture applications.

## 6.2 RF-Challenged Applications

A key advantage of CSS is its high sensitivity, which enables long-range operation. This sensitivity also enables long ranges in extremely challenging RF environments such as implantable and body worn devices.

*6.2.1 Smart Contact Lens.* Smart contact lens can measure vital indicators such as glucose, sodium and cholesterol in tears and enable long term unobtrusive real-time tracking of such vital parameters [39, 65]. Although progress has been made in miniaturization of the sensor IC, antenna and packaging, real time communication remains a bottleneck [34]. Radio communication is not applicable since it consumes orders of magnitude higher power than available on miniature batteries on contact lens form factor devices [23, 39, 47]. Since the environment is highly unfavorable to RF propagation, only extremely short range backscatter of tens of centimeters has been feasible using both custom hardware [39, 65] and standards-compliant Wi-Fi/BLE hardware [34]. Such short ranges limit applicability and renders backscatter less practical.

We leverage LoRa backscatter to demonstrate that a smart contact lens can communicate using CSS modulated backscatter at orders of magnitude larger distances than was feasible with prior approaches. We built a contact lens form factor antenna shown in Fig. 16(a). The antenna is a 1 cm diameter loop 30 AWG wire encapsulated between two soft contact lenses for biocompatibility and structural integrity. We immerse the antenna in contact lens solution and connect it to our COTS prototype.

We deploy the contact lens in a large atrium measuring 104 ft by 32 ft. We place the RF source and the receiver at the two ends of the room. We set the backscatter device to transmit LoRa packets at a 1 MHz offset and place the contact lens form factor antenna at the center of 12 ft by 10 ft grids. We log the reported RSSI value of the received packets at each of the tested locations. Fig. 16(b) plots the CDF of the received RSSI. Our results show reliable connectivity across the entire atrium, which is order of magnitude larger range than prior designs [34, 39].

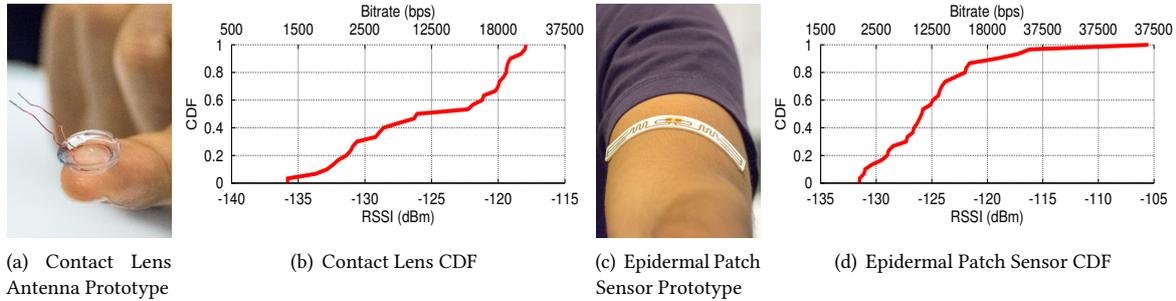

(a) Contact Lens Antenna Prototype  (b) Contact Lens CDF  (c) Epidermal Patch Sensor Prototype  (d) Epidermal Patch Sensor CDF

Fig. 16. **Smart Contact Lens and Flexible epidermal patch sensor.** We show the RSSI and data rate distribution across an entire 3,328 $ft^2$ (309 $m^2$) atrium.

### 6.2.2 Flexible Epidermal Patch Sensor.
Finally, flexible patch sensors can be used to monitor temperature, sweat, ECG and other vital signs in real time [24, 48]. These sensors are worn on the body and are in unfavorable RF environments since the antenna gets significantly detuned and degrades the link quality.

We use our system to prototype a form factor flexible epidermal patch sensor. To build our prototype, we use the sticker form factor antenna of an Alien swiggle RFID tag [2] and connect it to our COTS hardware. Using this approach, we can leverage the cost and size of RFID tags and create small, inexpensive and reliable epidermal patch sensors. We use a matching network to tune the impedance of the antenna when it's in contact with the human skin. We attach the sticker form factor antenna prototype on a subject's hand and test the device across the 3,328 $ft^2$ (309 $m^2$) atrium using the same deployment as the contact lens prototype with a 1 MHz backscatter offset. Fig. 16(d) plots the CDF of RSSIs for the received packets. Our results show that we can provide reliable connectivity with an RSSI greater than -132 dBm.

## 7 DISCUSSION AND CONCLUSION
We present the first backscatter system that can achieve the ranges required to enable wide-area communication. In this section, we outline avenues for future research.

*Achieving higher data rates.* Our current design achieves all the bit rates supported by LoRa. However an interesting research direction is to achieve higher data rates at the desired ranges, using either multiple frequency bands to concurrently transmit or multiple antenna designs.

*LoRa-independent design.* A reader might ask a very relevant question: does this technology depend on the wide adoption of LoRa? The answer is that if LoRa is successful, our technology can ride on top of its adoption. However the key component of our wide area backscatter design is the CSS modulation, which can provide the long ranges and high sensitivities independent of LoRa.

*RF Power harvesting.* LoRa backscatter is independent of how it is powered. We noted that our current design can be powered using very small solar cells, button cells as well as printed batteries, given its low power consumption. One can also explore research opportunities for long-range power harvesting [58] from the RF source.

*Networking LoRa backscatter devices.* The focus of this paper was to design and implement the LoRa backscatter physical layer. We believe that there are a number of research opportunities in networking 100-1000s of tiny devices that use LoRa backscatter communication modules in the vicinity of each other.